\input harvmac
\overfullrule=0pt
\input epsf.tex
\def\b{\bigskip}

\def\q{\quad}
\def\bu{$\bullet$ }

\Title{}
{\vbox{\centerline{ How does the Universe expand?
 }}}
\smallskip
\smallskip
\centerline{\bf Samir D. Mathur}
\smallskip
\bigskip

\centerline{\it Department of Physics}
\centerline{\it The Ohio State University}
\centerline{\it Columbus, OH 43210, USA}
\medskip
\centerline{mathur@mps.ohio-state.edu}
\bigskip

\medskip

\noindent

Quantization of gravity suggests that a finite region of space has a finite number
of degrees of freedom or `bits'. What happens to these bits when spacetime  expands,
as in cosmological evolution? Using 
gravity/field theory duality   we argue that bits `fuse together' when space expands.

\vskip 1.0 true in
[This essay received an "honorable mention"
        in the 2003 Essay Competition of the Gravity
        Research Foundation.]

\vskip 2.0 true in

\Date{}

A quantum cosmologist, arguing very na\"{\i}vely, finds the following
puzzle. He takes a spacelike region which has a proper volume $V$
at time $t$. Quantum gravity tells him that there is a cut-off at
planck length $l_p$, so he has $N={V/ l_p^3}$  `cells' or `bits'.
He thus expects the quantum theory to be described by a Hilbert
space with dimension $\sim e^N$. But the volume $V$ expands, and
for $t'>t$ seems to have $N'={V'/ l_p^3}>N$ bits. Since in quantum
mechanics the dimension of the Hilbert space must stay fixed,
 he concludes that no standard quantum theory can handle
cosmological expansion.
 
The simplest retort to this argument is -- there is {\it no}
cutoff at planck length; fourier modes of quantum fields have
wavelengths that go all the way down to $\lambda=0$.  But wait:
{\it this} assumption leads to trouble with {\it black holes}. In
Hawking's computation of radiation \ref\haw{S. Hawking: 1975,  {\it Comm. Math. Phys.} {\bf 43}, 
 199.} it is assumed that vacuum modes
with $\lambda<<\l_p$ exist; such modes dilate as they evolve near
the horizon and eventually become the radiation quanta. This
computation, if correct, gives information loss. Faced with this
failure of quantum mechanics we exclaim: But it is wrong to use
transplanckian modes so na\"{\i}vely! In a `correct' derivation of
radiation the  modes with $\lambda<<l_p$ are to be
replaced by {\it nonlocal } data; nonlocality occurs across
such large distances that information in the singularity ($r=0$)
is encoded in the radiation that is apparently leaving from
$r=R_{Schwarzschild}$.
 
Today most physicists would probably agree that in some sense
there are a finite number of degrees of freedom in a finite
region; in fact holography suggests that $N$ is even smaller,
given by the surface area of the region in planck units \ref\thooft{G. 't Hooft, gr-qc/9310026;  L. Susskind, J. Math. Phys. {\it
36} (1995) 6377.}. We are really up against the most basic question: What are the `bits'
making up spacetime, and how do these bits behave when spacetime
deforms, as for example in cosmological expansion? Without a
quantum description of spacetime we cannot understand why $\Lambda$ is finite (and small),  or what
determines the wavefunction of the Universe. Inflation expands a
planck volume by an enormous factor and derives $\delta\rho/\rho$
by freezing vacuum fluctuations; this makes it imperative to have
some insight into how degrees of freedom reshuffle when spacelike
slices  `stretch'.

To address this issue we look at a different system where we also see a `stretching' of space but where
we also have an {\it exact} quantum description of spacetime.

Maldecena's duality \ref\mal{J. Maldacena: 1998,
{\it Adv. Theor. Math. Phys.} {\bf 2},
231;  S. Gubser, I. Klebanov and A. Polyakov: 1998,
{\it Phys. Lett.} {\bf B428},
105;  E. Witten: 1998,
{\it Adv. Theor. Math. Phys.} {\bf 2},
253.} says that string theory (describing quantum gravity) on certain spacetimes has an exact
dual description as a {\it field theory}.   One case where this duality has been established in detail is the
`D1-D5 system' \ref\many{A. Strominger and C. Vafa: 1996, 
{\it Phys. Lett.} {\bf B379}, 
 99;  C. Callan and J. Maldacena: 1996, 
{\it Nucl. Phys.} {\bf B472}, 
 591;  S.R. Das and S.D. Mathur: 1996, 
{\it Nucl. Phys.} {\bf B478}, 
561.}\ref\lmtwo{
O.~Lunin and S.~D.~Mathur,
Commun.\ Math.\ Phys.\  {\bf 227}, 385 (2002)
[arXiv:hep-th/0103169].}\ref\lmfour{
O.~Lunin and S.~D.~Mathur,
Nucl.\ Phys.\ B {\bf 623}, 342 (2002)
[arXiv:hep-th/0109154].}. The field theory is described by an `effective string' which has winding number
$N$ around a circle of length $L$.  This string can be wound as $N$ separately closed loops (fig.1a), or
joined up into a single `multi-wound' string (fig.1b), or more generally have $m$ `component strings',
each of which  winds $n_i$ times before closing:
\eqn\one{\sum_{i=1}^m n_i=N.}
All these states have the same mass and charge, but from fig.1 we see that their dual gravity
descriptions have `throats' that `stretch' to different depths \lmfour .  We will shortly argue that each
of the $m$ component strings is a `bit', but we already see the moral: {\it When spacetime stretches, bits fuse
together.}

\bigskip
\bigskip

\epsfbox{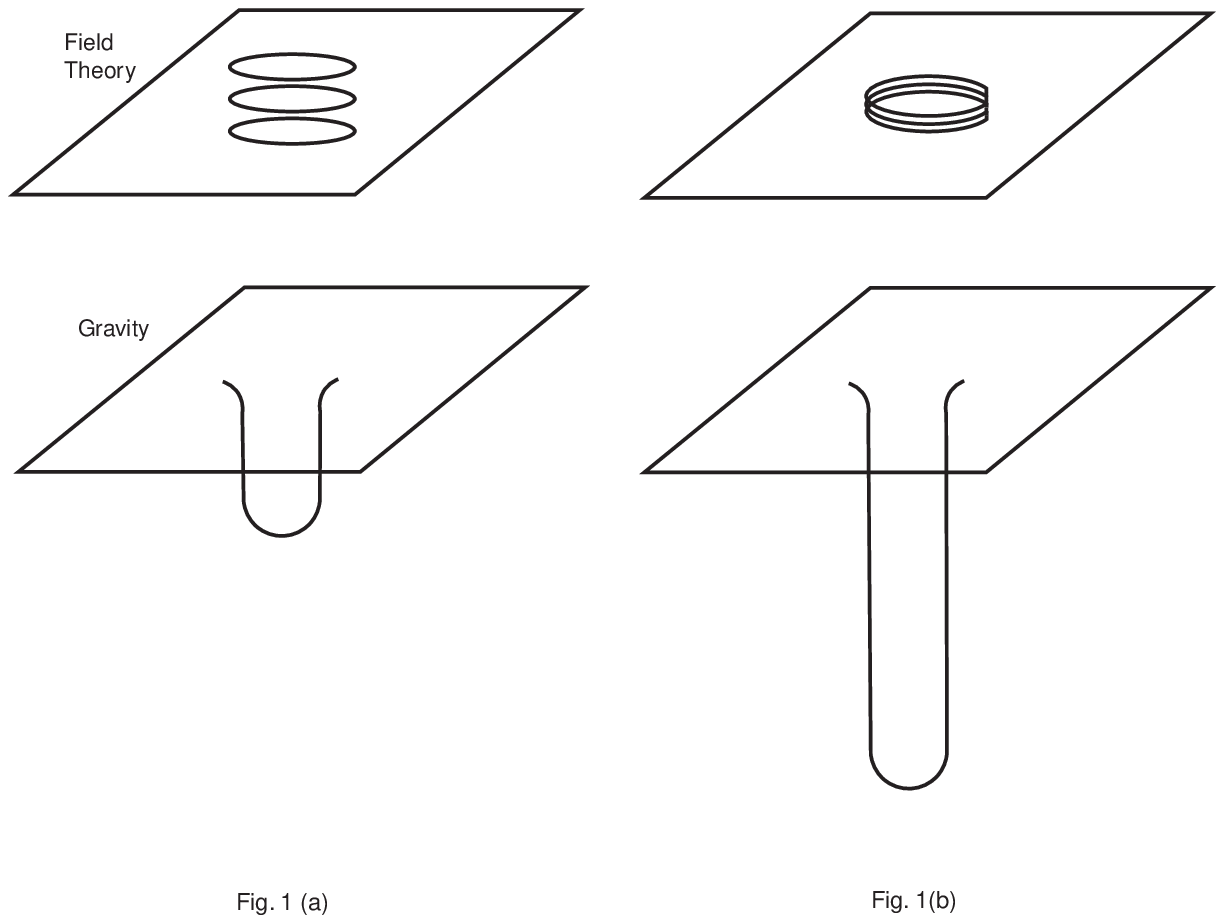}

\bigskip
\bigskip

To see why component strings are `bits' consider the dynamics in the field theory and gravity descriptions \lmfour .
A graviton incident on the effective string gets absorbed, its energy converted to a left moving
vibration (L) and a right moving vibration (R). In fig.1a the L and R excitations travel around the loop,
collide and exit the string after time $\Delta t_{string}=L/2$, while in fig.1(b) the string loop is $N$ times
longer and $\Delta t_{string}=NL/2$. In the {\it gravity} description the incident quantum just falls down
the throat and returns back up, in each case after a time 
\eqn\six{\Delta t_{gravity}=\Delta t_{string} ={L  \over 2}\langle n_i\rangle = {L\over 2} {N\over \langle m\rangle}  }

If we have {\it two} pairs of vibrations $(L,R)$ and $(L',R')$ in the field theory then they interact 
only if both pairs are carried by the {\it same} component string. This makes the effective interaction strength
$G_{eff}^{string}\sim 1/m$. 

In the dual geometries there is a compact direction of length $L'$ (not drawn). $L'$ is smaller for longer throats, and we
again find
\eqn\seven{G_{eff}^{gravity}={G\over L'}\sim {1\over m}}
($G$ is Newton's constant.)

When the number of quanta in the throat exceeds $\sim m$ we find that the gravitational backreaction becomes
order unity  and a horizon forms. In the dual field theory the presence of more than $m$ excitations
means that two or more excitations would be forced to live on  the same component string -- the
`bits' are all `used up'.

To summarize, longer throats are described by fewer component strings (bits), and each particle placed in the throat
corresponds to exciting one component string.

Let us now apply these lessons to cosmological expansion.  Imagine that the region marked in fig.2a
inflates as shown in fig.2b. A spherical wave sent in towards $r=0$ in the spacetime of fig.2a 
returns back in some time $\Delta t_1$. For the spacetime of fig.2b it would return after $\Delta
t_2>>\Delta t_1$. 
\b
\bigskip
\bigskip

\epsfbox{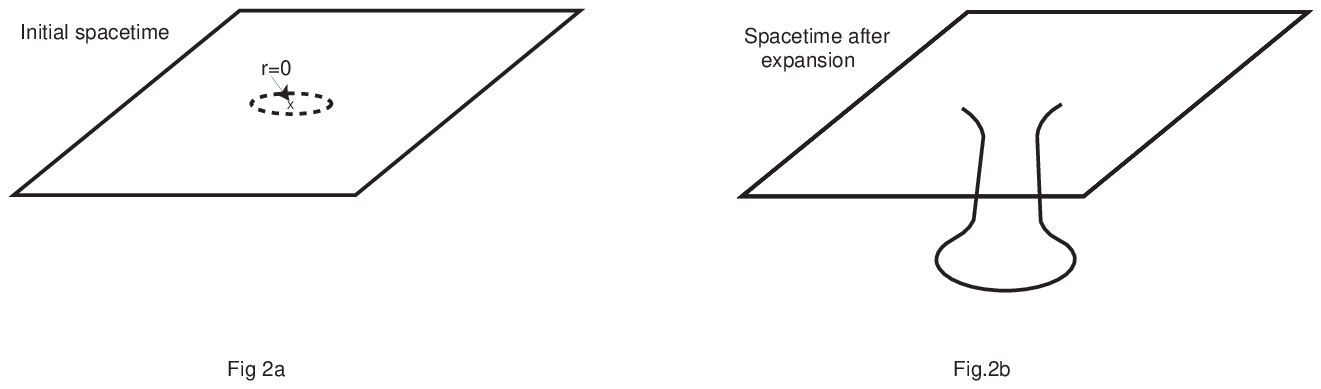}

\bigskip
\bigskip

In this case we do not know the map to a field theory dual, but based on intuition from the
exact duality studied above we expect the following:

\b

\bu  {\it Fusion of bits:}\q The marked region in fig.1a is described by
$N_i$ bits; these fuse together to give $N_f$ bits for the inflated region
in fig.2b with
\eqn\three{N_f~<~N_i}

\b

\bu {\it Evolution equation:}\q The D1-D5 states pictured in fig.1 are
stable, but if we break supersymmetry either by adding energy or by
modifying the action then we get an interaction Hamiltonian $H_{int}$
which cuts and joins loops \lmfour\
\eqn\four{\epsfbox{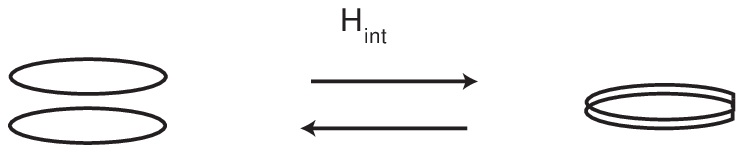} }
Such would be the  fundamental equation describing the evolution of
spacetime  at the quantum level. (The particles on the
spacetime are excitations of the bits, and these excitations get
rearranged when there is a merger of the bits they live on. )
Einstein's equations give only an effective low energy description which treats spacetime as an
infinitely stretchable smooth manifold. 
\b

\bu {\it Increase of effective coupling:}\q  When the space expands and bits fuse we expect that the effective coupling
$G_{eff}$ will increase. The simplest way to realize this would be to have a compact direction (as in the D1-D5 system) whose
length shrinks as the visible directions stretch.

\b

Physicists have long sought to describe spacetime in terms of discrete bits \ref\bousso{For recent work and
some past references see
 R.~Bousso,
JHEP {\bf 0011}, 038 (2000)
[arXiv:hep-th/0010252],
A.~Ashtekar,
arXiv:gr-qc/9302024.
}. A crucial
result of our analysis is that bits are `dynamic' objects that must join and split as spacetime deforms.  Near the cosmological
singularity ($t=0$)  space is minimally stretched, so $m$ will be large and the excitations representing different particles in
spacetime will be typically carried  by different bits.  (This would validate the proposal 
\ref\bkl{V.A. Belinski, I.M. Khalatnikov and E.M. Lifshitz, Adv. Phys. {\bf 19} (1970) 525, Adv. Phys. {\bf 31} (1982) 639.}
that there be a decoupling of dynamics between nearby points when $t\rightarrow 0$.)  At the other extreme, expansion
will stop when all bits have fused together ($m=1$). This is a very nonperturbative quantum gravity  effect, and
cannot be seen by solving the classical Einstein equations or  by performing a  semiclassical quantization of gravity. 

\listrefs

\end